\documentclass[twocolumn,
aps,
prx,
showpacs,
amsmath,
amssymb,
floatfix,
longbibliography,
superscriptaddress,
eprint]
{revtex4-1}

\usepackage{subfigure}

\usepackage{xcolor}
\definecolor{darkblue}{RGB}{0,0,150}
\definecolor{nightblue}{RGB}{0,0,100}

\usepackage{graphicx,mathtools,bm,bbm}
\usepackage{MnSymbol}
\usepackage[
colorlinks,
citecolor=darkblue,
linkcolor=darkblue,
urlcolor=nightblue]{hyperref}

\usepackage[english]{babel}
\usepackage[babel,kerning=true,spacing=true]{microtype}
\usepackage[utf8]{inputenc}

\usepackage{soul}

\usepackage{hyperref}
\hypersetup{
    colorlinks=true,
    linkcolor=blue,
    filecolor=magenta,      
    urlcolor=cyan,
    pdftitle={Overleaf Example},
    pdfpagemode=FullScreen,
    }

\newcommand{\vpd}{\vphantom{\dagger}}

\newcommand{\J}{\mathbf{J}}

\newcommand{\Tr}[1]{{\mathrm{Tr}\left[#1\right]}}

\begin{document}



\title{Efficiently preparing chiral states via fermionic cooling on bosonic quantum hardware}


    
\author{Gilad Kishony}\email{gilad.kishony@weizmann.ac.il}
\affiliation{Department of Condensed Matter Physics,
Weizmann Institute of Science,
Rehovot 76100, Israel}
\author{Mark S.~Rudner}
\affiliation{Department of Physics, University of Washington, Seattle, WA 98195-1560, USA}
\author{Erez Berg}
\affiliation{Department of Condensed Matter Physics,
Weizmann Institute of Science,
Rehovot 76100, Israel}
    
\begin{abstract}

Simulating many-body systems is one of the most promising applications of near-term quantum computers. An important open question is how to efficiently prepare the ground states of arbitrary fermionic Hamiltonians, especially those with nontrivial topology. Here, we propose an efficient protocol for preparing low-energy states of fermionic Hamiltonians on a noisy bosonic quantum simulator by adiabatic cooling using a simulated bath. We arrange the couplings such that the simulated system and bath together obtain a local fermionic description in which fermionic excitations can be extracted individually, via coherent hopping to the bath, rather than in pairs as would otherwise be required by fermion parity conservation. This approach thus achieves a linear (rather than quadratic) scaling of the cooling rate vs. excitation density at low densities. We show that certain topological phases such as the chiral (non-Abelian) phase of the Kitaev honeycomb model can be prepared efficiently using our protocol. Our protocol performs favorably in the presence of noise, making it suitable for execution on near-term quantum devices.

\end{abstract}

\maketitle

\section{Introduction}

Simulating quantum many-body systems, both in and out of equilibrium, is one of most promising applications of near-term quantum computers \cite{2012CiracZoller,Georgescu_2014}. In such tasks, quantum computers offer a vast advantage over their classical counterparts, by virtue of their ability to store the many-body wave functions and perform unitary gates 
that implement physical time evolution according to arbitrary Hamiltonians, at a polynomial cost in the system size. Such simulations may have practical utility in many fields, such as materials science, high-energy physics, and quantum chemistry. Preparation of a low-energy state of a many-body Hamiltonian is an important subroutine for these simulations. Doing this efficiently for arbitrary Hamiltonians is a highly non-trivial task, and is one of the major challenges in the field.

Of particular interest is the case where the ground state of the Hamiltonian is topologically non-trivial. Predicting the occurrence of topological states, e.g., in quantum spin systems or interacting electrons
in flat Chern bands, is notoriously difficult. In addition, such topological states are particularly challenging to prepare on quantum computers using conventional methods since, by definition, they cannot be connected to a product state by a finite-depth spatially-local unitary circuit \cite{Chen2010}. These unitary circuit algroithms include variational methods \cite{McClean_2016, Moll_2018, Tilly_2022}, adiabatic processes \cite{farhi2000quantum, Childs_2001, Aspuru_Guzik_2005, Albash_2018}, and effective imaginary time evolution \cite{Motta_2019, Lin2021Compressed,Jouzdani_2022} . In addition, protocols for cooling the system by coupling it to a simulated thermal bath have been proposed~\cite{Boykin_2002, Kaplan2017, Metcalf_2020, Polla_2021, Zaletel_2021,  RodriguezThesis, mathhies2022adibatic_demag,lloyd2024quasiparticle}. All these methods are expected to 
suffer a parametric reduction in performance
when preparing topological states, requiring
the circuit depth to increase with the system size to achieve a given accuracy.

A large class
of topologically non-trivial states can be prepared efficiently using dynamic circuits with
measurements and classical feedback \cite{Piroli2021,tantivasadakarn2022hierarchy}. The principle is illustrated by
the example of the toric code model~\cite{Kitaev2003}: the Hamiltonian is ``frustration--free,''
i.e., it consists of a sum of commuting terms that can be measured
simultaneously. After the measurement, the system can be brought into
its ground state by applying a unitary transformation designed to
remove the excitations that were detected by the measurement. This
method is not applicable for generic Hamiltonians away from the frustration-free
limit. 
Moreover, certain topological
states, such as chiral states,
do not have a frustration-free description. 

In this study, we tackle the problem of preparing chiral topological states, focusing on the well-known example of the chiral gapped phase of
the Kitaev spin liquid (KSL) model on the honeycomb lattice, which hosts non-Abelian excitations~\cite{kitaev2006anyons}. 
The KSL model can be mapped into a system of emergent Majorana fermions coupled to a static $\mathbb{Z}_2$ gauge field. 

\begin{figure}[b]
\begin{centering}
\includegraphics[width=(\textwidth-\columnsep)/2]{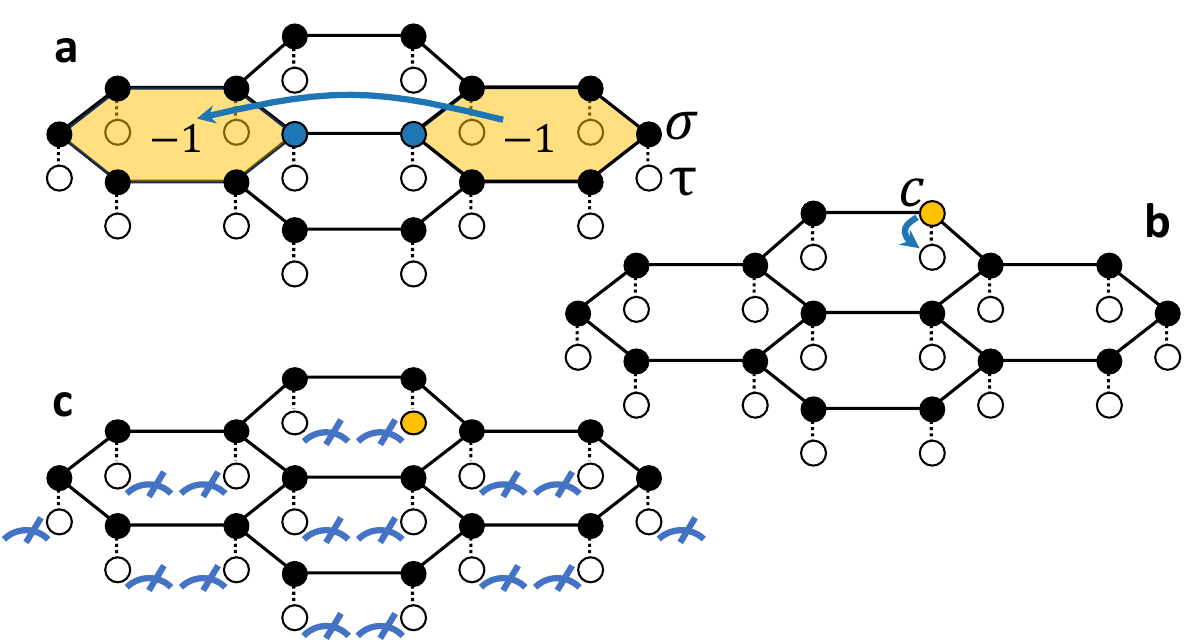}
\par\end{centering}
\caption{\textbf{Scheme for simulated cooling of a fermionic system using bosonic qubits.} The system's fermionic degrees of freedom (black) together with a fermionic ``bath'' (white) are encoded using bosonic qubits such that the two are charged under the same $\mathbb{Z}_2$ gauge field. \textbf{(a)} The system is brought to a flux-free state by measuring the flux operators and annihilating defects in pairs by applying a string of spin operators. \textbf{(b)} The system is evolved unitarily with the time-dependent Hamiltonian in Eq.~\eqref{eq:H_KSL}, allowing single fermionic excitations to hop coherently from the system to the bath. \textbf{(c)} At the end of the cooling cycle, the hopping between the system and bath sites is turned off. The location of the fermionic excitations in the bath is measured and the measurement outcomes are used to determine the Hamiltonian for the next cycle.}
\label{fig:protocol}
\end{figure}

The preparation scheme consists of two steps. 
In the first step, the system is prepared in the flux-free sector by measuring the flux operators through all plaquettes, and applying unitary operators that correct the fluxes that are detected~\cite{Dennis_2002}.
The second step is to prepare the fermionic ground state, whose Chern number is non-zero. The key idea is to introduce auxiliary ``bath'' degrees of freedom that simulate an effective fermionic bath,  coupled to the system's emergent fermions. 
Both species of fermions are coupled to the same $\mathbb{Z}_2$ gauge field, allowing single-fermion hopping between them. 
We provide an explicit encoding of the system and bath fermions into the physical bosonic qubits, generalizing the one-dimensional construction of Ref.~\cite{kishony2023gauged} to two dimensions. (By bosonic qubits, we mean that operators belonging to different qubits commute with each other. Qubits can be thought of as hard-core bosons.)
The bath fermions are then used to extract energy and entropy from the system, by performing simulated cyclic adiabatic cooling~\cite{mathhies2022adibatic_demag,kishony2023gauged} followed by measurements of the auxiliary qubits. The steps of the protocol are illustrated in Fig. \ref{fig:protocol}.

Our protocol prepares the ground state of the chiral phase parametrically more efficiently than other methods, such as adiabatic preparation~\cite{farhi2000quantum} or `naive' adiabatic cooling using a simulated bosonic bath~\cite{Kaplan2017,mathhies2022adibatic_demag}. These methods require a quantum circuit whose depth grows at least polynomially with the system size. In contrast, within our method, the energy density decreases exponentially with the number of cycles performed, and thus the total depth required to achieve a given accuracy in the total ground state energy grows only logarithmically with the system size. In the presence of decoherence, our protocol reaches a steady state whose energy density is proportional to the noise rate, which is parametrically lower than bosonic cooling protocols (where the energy density of the steady state scales as the square root of the noise rate~\cite{mathhies2022adibatic_demag}). 

The protocol presented here can be used to efficiently prepare the ground state of a generic (interacting) fermionic model in 2D. This is done by artificially introducing a $\mathbb{Z}_2$ gauge field similar to the physical gauge field of the KSL model in order to facilitate the fermionization scheme for the system and the bath. The presence of an effective fermionic bath allows to remove single fermion excitations, while keeping the coupling between the system and auxiliary qubits spatially local.


\section{Results}
\subsection{Preparing a chiral spin liquid}
\label{sec: preparing a chiral spin liquid}

\subsubsection{Model}

In this section we explain how 
to prepare the ground state of the KSL model in the phase where the emergent fermions carry a non-trivial Chern number, 
\emph{as efficiently as in a topologically trivial phase}. The same principle can be applied to arbitrary interacting fermionic Hamiltonians, as we discuss below.

\begin{figure}
\begin{centering}
\includegraphics[width=(\textwidth-\columnsep)/2]{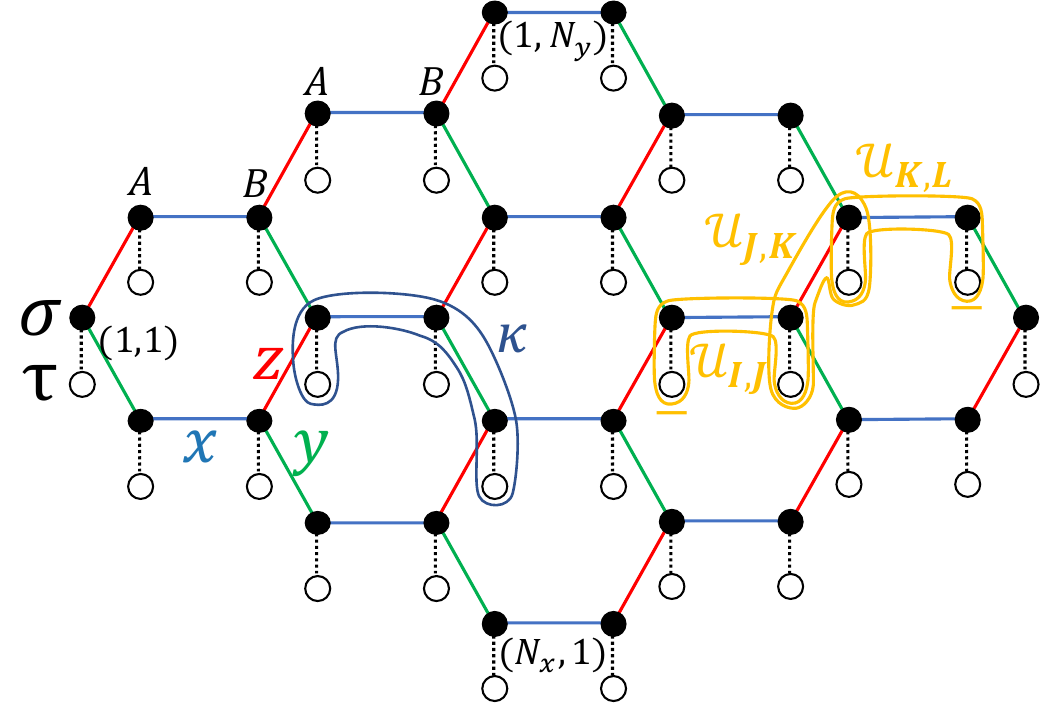}
\par\end{centering}
\caption{\textbf{The generalized KSL model.} Each unit cell consists of two system spins 
(black) and two bath 
spins (white). The bonds are colored according to orientation: x-blue, y-green, z-red. The five-spin term $\kappa$ drives the system into the chiral phase. A product of the unitaries $\mathcal{U}_{\mathbf{I},\mathbf{J}}$ along a path removes a pair of fermionic excitations from the bath. Instead of applying this non-local operation to the state, any subsequent operator acting on the system is conjugated by it.}
\label{fig. ksl}
\end{figure}

The KSL model consists of spin-$\frac{1}{2}$ degrees of freedom on a honeycomb lattice with strongly anisotropic exchange interactions~\cite{kitaev2006anyons,winter2017models,hermanns2018physics}. The Hamiltonian is written as:
\begin{align}
H_{\text{KSL}}=&-\sum_{\alpha\in\{x,y,z\}}\mathcal{J}_{\alpha}\sum_{\langle\mathbf{I},\mathbf{J}\rangle\in \alpha\text{-bonds}}\sigma_{\mathbf{I}}^{\alpha}\sigma_{\mathbf{J}}^{\alpha}
\nonumber\\
&-\kappa\sum_{\langle\langle\mathbf{J},\mathbf{K},\mathbf{L}\rangle\rangle}\sigma_{\mathbf{J}}^{x}\sigma_{\mathbf{K}}^{y}\sigma_{\mathbf{L}}^{z}.
\label{eq: KSL}
\end{align}
Here, $\mathbf{J} = (i,j,s)$ denotes the sites of the honeycomb lattice, where $i \in \{1,\dots,N_x\}$ and $j \in \{1,\dots,N_y\}$ label the unit cell and $s\in \{A,B\}$ labels the two sublattices, 
and $\sigma_{\mathbf{J}}^\alpha$ denotes the $\alpha=x,y,z$ Pauli matrix at site $\mathbf{J}$. 
The bonds of the lattice are partitioned into three sets -- $x$, $y$, and $z$ bonds --according to their geometric orientations (see Fig.~\ref{fig. ksl}).
We denote the corresponding Kitaev anisotropic exchange couplings by $\mathcal{J}_x, \mathcal{J}_y,$ and $\mathcal{J}_z$, respectively. 
The $\kappa$ term is a three-spin interaction acting on three neighboring sites denoted by $\langle\langle \mathbf{J},\mathbf{K},\mathbf{L}\rangle \rangle$, 
such that both $\mathbf{J}$ and $\mathbf{K}$ are nearest neighbors of site $\mathbf{L}$. 
The $\kappa$ term breaks time reversal symmetry, and opens a gap in the bulk 
that drives the system into the chiral non-Abelian phase~\cite{kitaev2006anyons}. 

In order to facilitate an efficient protocol to prepare the ground state of \eqref{eq: KSL}, we introduce a modified KSL Hamiltonian, $\widetilde{H}_{\text{KSL}}$, that includes both a system ($\sigma$) and a auxiliary ($\tau$) spin at every site of the honeycomb lattice. The auxiliary spins are used to cool the system down to its ground state. 
The Hamiltonian is illustrated in Fig.~\ref{fig. ksl}, and is composed of a sum of two terms:
\begin{align}
H(t)=\widetilde{H}_{\text{KSL}}+H_{{\rm c}}(t),
\label{eq:H_KSL}
\end{align}
where $\widetilde{H}_{\text{KSL}}$ is the modified KSL Hamiltonian:
\begin{align}
\widetilde{H}_{\text{KSL}}=&-\sum_{\alpha\in\{x,y,z\}}\mathcal{J}_{\alpha}\sum_{\langle \mathbf{I},\mathbf{J}\rangle\in \alpha\text{-bonds}}\sigma_{\mathbf{I}}^{\alpha}\sigma_{\mathbf{J}}^{\alpha}\tau_{\mathbf{I}}^{z}\tau_{\mathbf{J}}^{z}\nonumber\\
&-\kappa\sum_{\langle\langle \mathbf{J},\mathbf{K},\mathbf{L}\rangle\rangle}\sigma_{\mathbf{J}}^x\tau_{\mathbf{J}}^z\sigma_{\mathbf{K}}^y\tau_{\mathbf{K}}^z\sigma_{\mathbf{L}}^z.
\label{eq: KSL with spins}
\end{align}
The $\tau^z_{\bf J}$ operators all commute with $\widetilde{H}_{\rm{KSL}}$. Clearly, $\widetilde{H}_{\text{KSL}}$ is identical to  $H_{\text{KSL}}$ in Eq. \eqref{eq: KSL} in the sector $\tau^z_{\mathbf{J}}=+1$. In fact, the two Hamiltonians are related by a unitary transformation for any state of the $\tau$ spins with even parity, $\prod_{\mathbf{J}} \tau^z_{\mathbf{J}}=1$. 
To see this, 
we note that 
all of the unitary operators 
$\mathcal{U}_{\mathbf{I},\mathbf{J}} = \tau^x_{\mathbf{I}}\sigma^\alpha_{\mathbf{I}}\sigma^\alpha_{\mathbf{J}}\tau^x_{\mathbf{J}}$, for $\langle \mathbf{I},\mathbf{J}\rangle\in \alpha\text{-bonds}$, commute with $\widetilde{H}_{\rm{KSL}}$ and anticommute individually with $\tau^z_{\mathbf{I}}$ and $\tau^z_{\mathbf{J}}$. The transformation from a sector with a given $\{\tau^z_\mathbf{J}\}$ of even parity to the sector \{$\tau^z_\mathbf{J}=1$\} can be constructed by annihilating pairs of $\tau^z_\mathbf{K},\tau^z_\mathbf{L}=-1$ using a product of the unitaries $\mathcal{U}_{\mathbf{I},\mathbf{J}}$ along a path connecting the points $\mathbf{K}, \mathbf{L}$. Thus, preparing the ground state of $\widetilde{H}_{\rm{KSL}}$ in any known even-parity sector of $\{\tau^z_{\mathbf{J}}\}$ is equivalent to preparing the ground state of the original KSL Hamiltonian, Eq.~\eqref{eq: KSL}. 

The ``control'' Hamiltonian $H_{{\rm c}}(t)$, which is used to cool into the ground state of $\widetilde{H}_{\text{KSL}}$, is chosen as a time-dependent effective Zeeman field acting on $\tau_{\mathbf{J}}$ with $\hat{x}$ and $\hat{z}$ components $g_{\J}(t)$ and $B_{\mathbf{J}}(t)$, respectively, 
 \begin{align}
 \label{eq: control hamiltonian}
H_{{\rm c}}(t)=-\sum_{\mathbf{J}}B_{\mathbf{J}}(t)\tau_{\mathbf{J}}^{z}-\sum_{\mathbf{J}}g_{\J}(t)\tau_{\mathbf{J}}^{x}.
 \end{align}
The time dependence of $B_{\mathbf{J}}(t)$ and $g_{\J}(t)$ and the protocol used for cooling are described below.

\subsubsection{Fermionization}

It is useful to fermionize the spins $\sigma$ and $\tau$ by introducing a set of six Majorana operators $\{c_{\mathbf{J}}^{x},c_{\mathbf{J}}^{y},c_{\mathbf{J}}^{z},b_{\mathbf{J}}^{x},b_{\mathbf{J}}^{y},b_{\mathbf{J}}^{z}\}$ per site, subject to the constraint
\begin{equation}
c_{\mathbf{J}}^{x}c_{\mathbf{J}}^{y}c_{\mathbf{J}}^{z}b_{\mathbf{J}}^{x}b_{\mathbf{J}}^{y}b_{\mathbf{J}}^{z}=i.
\label{eq:constraint}
\end{equation}

The spin operators are related to the fermionic ones by: 
\begin{equation}
\sigma_{\mathbf{J}}^{\alpha}  =-\frac{i}{2}\sum_{\beta,\gamma}\varepsilon^{\alpha\beta\gamma}b_{\mathbf{J}}^{\beta}b_{\mathbf{J}}^{\gamma},\quad
\tau_{\mathbf{J}}^{\alpha}  =-\frac{i}{2}\sum_{\beta,\gamma}\varepsilon^{\alpha\beta\gamma}c_{\mathbf{J}}^{\beta}c_{\mathbf{J}}^{\gamma},
\label{eq: mapping spins to fermions}
\end{equation}
where $\alpha,\beta,\gamma \in \{x,y,z\}$ and $\varepsilon^{\alpha\beta\gamma}$ is the totally antisymmetric
tensor. This transformation of the spins to fermions is analogous to the one used by Kitaev~\cite{kitaev2006anyons}. Note, however, that we have fermionized the $\sigma$ and $\tau$ spins together, and as a result, there is a single constraint per site containing both a $\sigma$ and a $\tau$ spin. 
For later use, we note that, using Eqs.~(\ref{eq:constraint}) and (\ref{eq: mapping spins to fermions}), we can write
\begin{equation}
\label{eq:tausigma_cb} \tau_{\mathbf{J}}^{z}\sigma_{\mathbf{J}}^{\alpha}=-ic_{\mathbf{J}}^{z}b_{\mathbf{J}}^{\alpha}.
\end{equation}

Using this mapping, we express 
Hamiltonian \eqref{eq:H_KSL} as 
\begin{align}
H(t)=&\sum_{\alpha\in\{x,y,z\}}\mathcal{J}_{\alpha}\sum_{\langle \mathbf{I},\mathbf{J}\rangle\in \alpha\text{-bonds}}u_{\mathbf{I},\mathbf{J}}ic_{\mathbf{I}}^{z}c_{\mathbf{J}}^{z}\nonumber\\
&-i\kappa\sum_{\langle\langle \mathbf{J},\mathbf{K},\mathbf{L}\rangle\rangle}u_{\mathbf{J},\mathbf{L}}u_{\mathbf{L},\mathbf{K}}c_{\mathbf{J}}^{z}c_{\mathbf{K}}^{z}\nonumber\\
&-\sum_{\mathbf{J}}B_{\mathbf{J}}(t)ic_{\mathbf{J}}^{y}c_{\mathbf{J}}^{x}-\sum_{\mathbf{J}}g_{\J}(t)ic_{\mathbf{J}}^{z}c_{\mathbf{J}}^{y},
\label{eq:H_KSL_fermions}
\end{align}
where we identify the set of conserved $\mathbb{Z}_2$ gauge fields
\begin{align}
\left\{u_{\mathbf{I},\mathbf{J}}=ib_{\mathbf{I}}^{\alpha}b_{\mathbf{J}}^{\alpha}|\langle \mathbf{I},\mathbf{J}\rangle\in \alpha\text{-bonds},\alpha\in\{x,y,z\}\right\}.
\end{align}
The gauge-invariant flux $W_{i,j}$ of the $\mathbb{Z}_2$ gauge field is defined on each hexagonal plaquette that corresponds to the unit cell $(i,j)$, 
as the product of $u_{\mathbf{I},\mathbf{J}}$ on its edges. In terms of the spin degrees of freedom, this can be written as
\begin{align}
W_{i,j} = \sigma^x_{(i,j,B)}\sigma^z_{(i+1,j,A)}\sigma^y_{(i+1,j,B)}\nonumber\\
\times\sigma^x_{(i+1,j+1,A)}\sigma^z_{(i,j+1,B)}\sigma^y_{(i,j+1,A)}.
\label{eq: fluxes in terms of spins}
\end{align}

The fermionic Hamiltonian~\eqref{eq:H_KSL_fermions} with $g_{\J}=0, B_{\mathbf{J}}=0$ is identical to the Hamiltonian found by fermionizing the 
original KSL model~\cite{kitaev2006anyons}, with $\{c_{\mathbf{J}}^{z}\}$ playing the role of the ``system fermions.''
The remaining $\{c^x_{\bf J}\}$ and $\{c^y_{\bf J}\}$ operators describe the fermionic modes of the bath, 
while the $\{b^\alpha_{\bf J}\}$ operators introduced above Eq.~(\ref{eq:constraint}) only participate through their roles in setting the values of the conserved $\mathbb{Z}_2$ gauge fields $\{u_{{\bf I}, {\bf J}}\}$.
The term multiplied by $g_{\bf J}$ in Eq.~(\ref{eq:H_KSL_fermions}) allows fermions to hop between the system and the bath, 
while the term multiplied by $B_{\bf J}$ acts on the bath fermions alone. 
Our aim is to prepare the system fermions $\{c_{\mathbf{J}}^{z}\}$ in the ground state of the Hamiltonian \eqref{eq:H_KSL_fermions} with $g_{\J}=0$ and no $\mathbb{Z}_2$ fluxes, i.e., $W_{i,j}=1$ (equivalent to setting $u_{\mathbf{I},\mathbf{J}}=1$, up to a $\mathbb{Z}_2$ gauge transformation).

\subsubsection{Preparation Protocol}
We now present the protocol to prepare the ground state of $\widetilde{H}_{\text{KSL}}$, starting from an arbitrary initial state. Similarly to Refs. \cite{mathhies2022adibatic_demag,kishony2023gauged}, the protocol consists of cycles which are repeated until convergence to the ground state is achieved. 
Since the protocol is intended to be implemented on bosonic qubits, in this section we return to the description in terms of the $\sigma$ and $\tau$ spins.

In each cycle, the $\sigma$ and $\tau$ spins are evolved unitarily for a time period $T$, with a time-dependent Hamiltonian~\eqref{eq:H_KSL} designed to decrease the expectation value of $\widetilde{H}_{\text{KSL}}$. The unitary evolution is followed by a projective measurement of the $\tau$ spins in the $z$ basis. The measurement outcomes are used to determine the Hamiltonian of the next cycle. 

\begin{figure}
\begin{centering}
\includegraphics[width=(\textwidth-\columnsep)/3]{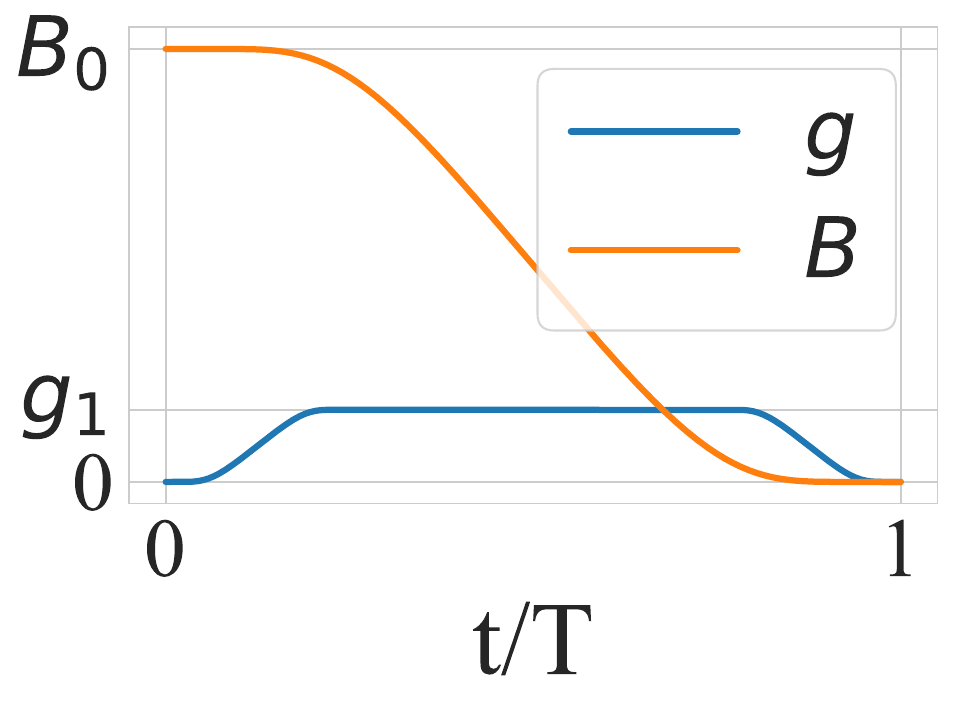}
\par\end{centering}
\caption{\textbf{Adiabatic evolution of the time dependent couplings $g$ and $B$.} The time dependence of the couplings $g$ and $B$ of the control Hamiltonian [Eq.~(\ref{eq: control hamiltonian})]. Explicit expressions are given in Methods subsection ``Smooth evolution of time dependent couplings''.}
\label{fig. g and B}
\end{figure}

In the beginning of the $n$th cycle, the $\tau$ spins are assumed to be in an eigenstate of $\tau^z_{\J}$ with known eigenvalues, which we denote by $\tau_{\J}(t_n)=\pm 1$. Before the first cycle, the system is initialized in a state where $\tau^z_{\mathbf{J}}=+1$ for all $\mathbf{J}$. We then perform the following operations:


\begin{enumerate}
    \item
    The $\sigma$ spins are brought to a flux-free state, $W_{i,j} = 1$. This can be done by measuring all the flux operators (which are mutually commuting), and annihilating the detected fluxes in pairs by applying appropriate unitary operators~\cite{Kalinowski_2023} (see Methods subsection ``Removing fluxes in 2D'' 
    for details). 

    \item  
    The system is evolved from time $t_n$ to $t_{n+1} = t_n + T$ with the Hamiltonian $H(t-t_n)$, where $H(t)$ is given in Eq.~\eqref{eq:H_KSL}.
    Specifically, we set the effective Zeeman fields to $B_{\mathbf{J}}(t) = B(t-t_n)\tau_{\mathbf{J}}(t_n)$  and the ``system-bath couplings'' to $g_{\mathbf{J}}(t) = g(t-t_n)$, with the functions $g(t)$, $B(t)$ as shown in Fig. \ref{fig. g and B} and given explicitly in Methods subsection ``Smooth evolution of time dependent couplings''.

    \item 
    The $\tau^z_{\J}$ operators are measured, giving new values $\{\tau_{\J}(t_{n+1})\}$. A new cycle then begins from step b.  
\end{enumerate}

Importantly, the flux operators $W_{i,j}$ commute with $H(t)$.
Ideally, they therefore retain their initial values, $W_{i,j}=1$. In the presence of decoherence on a real quantum simulator, $W_{i,j}$ may flip during the cycle. This may require returning to step 1 between cycles, where the flux operators are measured again and unitary operators are applied to correct flipped fluxes, as discussed in Sec.~``Performance with noise'' 
and in Methods subsection ``Removing fluxes in 2D''.

The idea behind this protocol is that, in analogy to adiabatic demagnetization, the coupling between the $\sigma$ and $\tau$ spins cools the system towards the ground state of $\widetilde{H}_{\text{KSL}}$. 
The $\tau$ spins are initially polarized in a direction parallel to the effective Zeeman fields acting on them. 
Adiabatically sweeping the Zeeman fields downward while the coupling $g$ is non-zero tends to transfer energy into the $\tau$ spins, decreasing the expectation value of $\widetilde{H}_{\text{KSL}}$. This picture guides the choice of the protocol parameters, $B_0$, $g_1$ and $T$: $B_0$ should be chosen to be large enough compared to the gap $E_{\text{gap}}$ between the ground state and the lowest excited state of $\widetilde{H}_{\text{KSL}}$, such that sweeping the Zeeman field during the protocol brings the excitations in the system to resonance with the $\tau$ spins. The maximum coupling $g_1$ controls the rate of the cooling. $T$ is chosen such that the time evolution is adiabatic with respect to the system's gap, which sets the maximum allowed values of $B_0$ and $g_1$: these are chosen such that $\text{max}(|B_0|, |g_1|)/T\ll E_{\text{gap}}^2$. 

The operation of the cooling protocol is particularly transparent in the fermionic representation, Eq. (\ref{eq:H_KSL_fermions}). It is straightforward to check that in the beginning of each cycle, when $g_{\J}=0$, the bath fermions $c^x$ and $c^y$ are decoupled from the system fermions, $c^z$, and are initialized in the ground state of the bath Zeeman Hamiltonian. Sweeping $|B_{\J}(t)|$ downwards induces level crossings between states of the system ($c^z$) and bath fermions, and when $g_{\J}\ne 0$, excitations are transferred from the system to the bath. 

\subsubsection{Performance analysis}

The fact that the system and bath can be described as emergent fermions which are coupled to the same gauge field means that a single fermionic excitation can be transferred coherently to the bath. This dramatically accelerates the cooling process compared to simpler ``bosonic'' adiabatic cooling algorithms of the type described in Refs. \cite{Boykin_2002, Kaplan2017, Metcalf_2020, Polla_2021, Zaletel_2021,  RodriguezThesis, mathhies2022adibatic_demag,lloyd2024quasiparticle}. In these protocols, only a gauge-neutral {\it pair} of fermionic excitations can be transferred from the system to the bath, slowing down the cooling process as the ground state is approached. 

Specifically, the performance of our protocol can be understood from simple considerations. The argument 
mirrors that given in Ref.~\cite{kishony2023gauged} for ``gauged cooling'' of a one-dimensional quantum Ising model. 
Since fermionic excitations in the system can independently hop into the bath, each excitation has a finite probability to be removed by the bath in each cycle. 
The cooling rate is therefore proportional to the energy density of $\widetilde{H}_{\text{KSL}}$. 
Approximating the cooling process as a continuous time evolution, we obtain a rate equation for the energy density, $\varepsilon(t)$:
\begin{equation}
    \dot{\varepsilon}(t) = -C \left[\varepsilon(t)-\varepsilon_{\rm s}\right],
    \label{eq: rate}
\end{equation}
where $C>0$ is the cooling rate, and $\varepsilon_{\rm s}$ is the steady state energy density. 
The quantities $C$ and $\varepsilon_{\rm s}$ depend on the protocol parameters, such as $T$, $B_0$ and $g_1$, and in a realistic noisy quantum simulator, also on the noise rate; in the perfectly adiabatic, noiseless case, $\varepsilon_{\rm s}\rightarrow \varepsilon_0$, the ground state energy density. Solving Eq. \eqref{eq: rate}, we obtain $\varepsilon(t) = \varepsilon_{\rm s} + [\varepsilon(0)-\varepsilon_{\rm s}]e^{-C t}$.
In contrast, in a simple adiabatic cooling protocol where only pairs of fermionic excitations can be transferred to the bath, $\dot{\varepsilon} \propto -(\varepsilon-\varepsilon_{\rm s})^2$, resulting in a power-law approach to the steady state, $\varepsilon-\varepsilon_{\rm s} \sim 1/t$. 

The fact that the fermionic Hamiltonian (\ref{eq:H_KSL_fermions}) is quadratic allows us to analyze the cooling dynamics in a basis in which the excitations are decoupled. Moreover, within the flux-free sector, and assuming that the system is prepared initially such that $\tau^z_{\mathbf{J}}=1$ and hence the signs of the $B_{\mathbf{J}}$'s are spatially uniform, the Hamiltonian can be brought into a translationally invariant form by a unitary transformation. 
This allows an explicit analysis of the performance of the protocol in momentum space, which we present in Supplementary Note 1, Supplementary Fig.~1 and Supplementary Fig.~2. 
The analysis verifies the exponential convergence to the steady state with the number of cycles performed, as anticipated above. Moreover, we find that in the ideal (noiseless) case, the steady state energy $\varepsilon_{\rm s}$ depends only on the adiabaticity of the protocol with respect to the system's gap, $E_{\text{gap}}$, whereas the cooling rate $C$ depends also on the the adiabaticity with respect to the avoided crossing gaps encountered during the time evolution with $H(t)$. In particular, in the limit where $T$ is large compared to $1/E_{\text{gap}}$, the ground state can be approached with an exponential accuracy.

While these results are demonstrated for the solvable model $\widetilde{H}_{\text{KSL}}$, which can be mapped to free fermions, we expect them to hold more generally. For example, the performance is not expected to be parametrically affected if quartic interactions between the system fermions are added to Eq.~(\ref{eq: KSL with spins}). In the presence of interactions, excitations retain a finite overlap with single fermions since the non-interacting model is gapped. The application of our protocol to general interacting fermionic Hamiltonians is discussed further in Sec.~``Extension to arbitrary fermionic Hamiltonians''.




 \subsection{Numerical simulations}
 \label{sec: numerics}
 \subsubsection{Performance without noise}
\label{sec: without noise}

We simulate the fermion cooling algorithm numerically with and without noise using the efficient procedure described in Methods subsection ``Single-particle density matrix simulation of free
fermions'' 
which relies on the model being one of free fermions. In all simulations we choose $\mathcal{J}_x=\mathcal{J}_y=\mathcal{J}_z=\mathcal{J}=1$, $\kappa=1$, in the chiral phase,
 and use protocol parameters $B_0=7$, $g_1=0.5$. Periodic boundary conditions are used unless otherwise stated.

 Setting periodic boundary conditions allows us to simulate this translation invariant model in $k$-space point by point. We cool a system of size $85\times85$ unit cells in the absence of noise starting from a completely random state. Here, it is assumed that the system has been initialized to the flux-free sector prior to the action of the cooling protocol, which leaves the gauge fields fixed. 
 Time evolution is performed 
 in each momentum sector using an ordinary differential equation solver with a relative tolerance of $10^{-6}$ such that the deviation of the steady state from the ground state is expected to derive predominantly from diabatic transitions.
 
 In the ground state, the Chern number of the $c^z_{\J}$ fermions is equal to 1. To probe the convergence to the ground state, we define the quantity $\nu^l$:
\begin{align}
\label{eq: chern number}
\nu^{l}=\frac{1}{2\pi i}\int \Tr{R^l\left[\partial_{k_x}R^l,\partial_{k_y}R^l\right]}dk_xdk_y,
\end{align}
 where $R^{l}$ is the single particle density matrix of either the system ($l=\rm{sys}$) or the bath ($l=\rm{bath}$) fermions. Specifically, $R^{\rm{sys}}_{s,s'}(\bm{k})=\left\langle c_{(\bm{k},s)}^{z\dagger} c_{(\bm{k},s')}^{z\vpd}\right\rangle$, with an analogous expression for $R^{\rm{bath}}$. 
 Here, $c_{(\bm{k},s)}^{z}$ is the Fourier transform of $c_{\J}^z = c_{(i, j ,s)}^z$ on sublattice $s$ with wavevector $\bm{k}$.  
 In a (pure) Slater determinant state, the quantity $\nu^{l}$ becomes the integrated Chern density. We calculate this value by performing a discrete integral in momentum space.
 
\begin{figure}
\begin{centering}
\includegraphics[width=(\textwidth-\columnsep)/2]{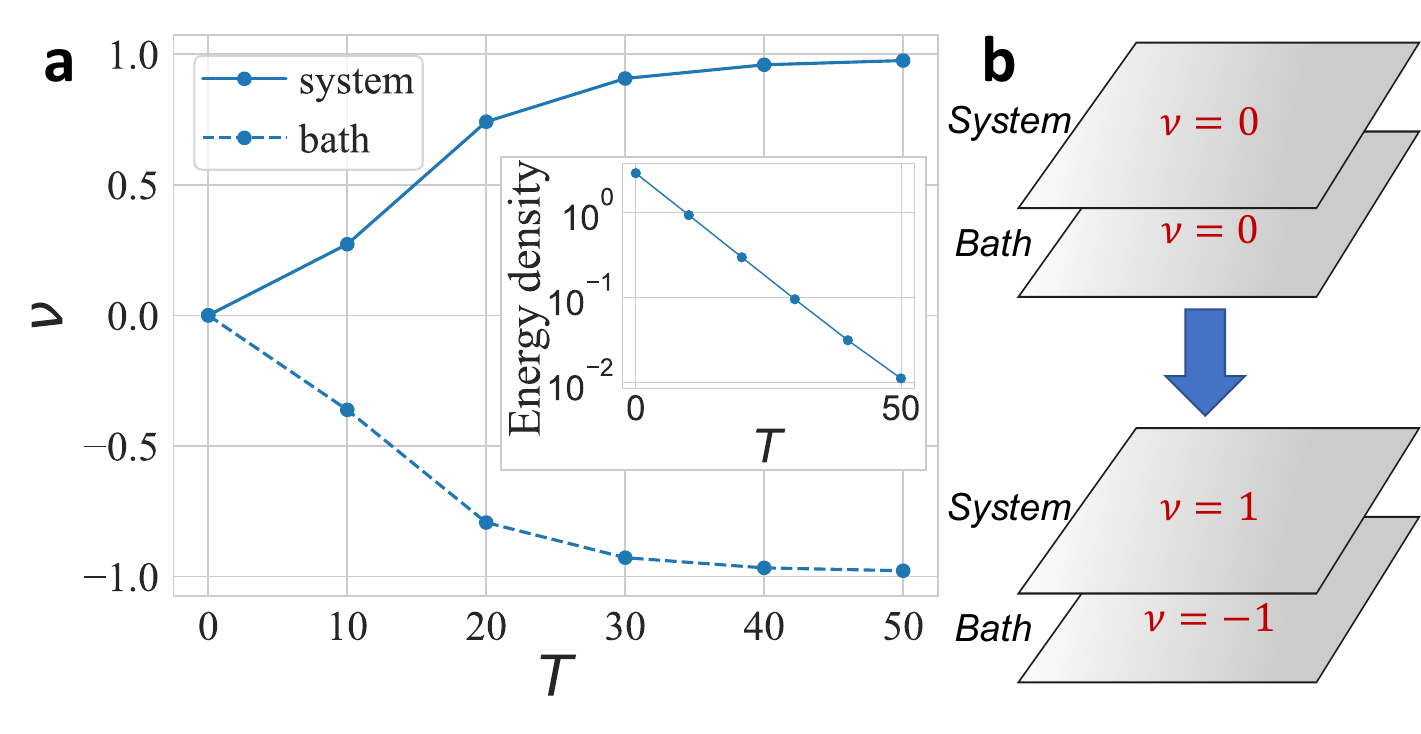}
\par\end{centering}
\caption{\textbf{A single cooling cycle performed on the KSL system.} A KSL system of size $85\times85$ with $\mathcal{J}=1,\kappa=1$ starting from a randomly initialized state in the flux-free sector is cooled for a single cycle using protocol parameters $g_1=0.5$, $B_0=7$. \textbf{(a)} A finite size version of the spectral Chern number \eqref{eq: chern number} of the system (solid lines) and the bath (dashed lines) before performing the reset is shown as a function of the cycle duration $T$. The inset shows the energy density vs. $T$. In the adiabatic limit, the system approaches the ground state, the Chern number of the system approaches $1$ and that of the bath approaches $-1$ in a single cycle as illustrated in \textbf{(b)}.}
\label{fig. KSL energy and chern number}
\end{figure}

 Fig.~\ref{fig. KSL energy and chern number} shows $\nu^{\rm{sys}}$ (solid lines) and $\nu^{\rm{bath}}$ (dashed lines) at the end of a single cooling cycle before performing the measurement of the $\tau$ spins, as a function of the 
 sweep time, $T$. As the 
 sweep duration is increased, the Chern number of the system approaches $1$ and that of the bath approaches $-1$, even after a single cycle. The fact that our protocol succeeds in preparing a fermionic 
 chiral ground state starting from a 
 topologically trivial state
 by a finite depth local unitary circuit is due to 
 the chiral state possessing ``invertible'' topological order; i.e., stacking two systems with opposite Chern numbers yields a trivial state. 
 Thus, the topological state of the system's fermions and its inverse (in the bath)
 can be prepared starting from a trivial state. Note, however, that overall, the spin system is in a {non-invertible} topologically ordered state, since the chiral phase of the KSL has fractional (non-Abelian) statistics. 

The inset of the Fig.~\ref{fig. KSL energy and chern number} shows the energy density after a single cooling cycle as a function of the duration $T$. As the cycle duration is increased, the system approaches the ground state energy in a single cycle. The deviation from the ground state energy decreases exponentially with increasing $T$.


\subsubsection{Performance with noise}
\label{sec: with noise}

In this section, we numerically test the performance of the proposed cooling algorithm in the presence of noise.
For simplicity, the noise is modeled as a uniform depolarizing channel acting on all $\sigma$ and $\tau$ qubits 
at some rate $\zeta$ of errors per cooling cycle per qubit. This noise is simulated stochastically as described in Methods subsection ``Single-particle density matrix simulation of free
fermions''.

Importantly, 
the fluxes of the gauge field may be excited by the action of noise on the system. In order to overcome this, one should periodically measure and fix these fluxes, while performing our proposed algorithm for cooling the excitations of the fermions $\{c_{\bf J}^z\}$. Fixing the observed fluxes can be done in an analogous way to the action of error correction in the surface code~\cite{Dennis_2002} by annihilating them in pairs, as explained in Methods subsection ``Removing fluxes in 2D''. 
When flux excitations are removed (by applying local spin operations), fermionic excitations may be created, and these are then cooled by subsequent cycles of the cooling protocol. Using the protocol, the energy density of the steady state should be proportional to the noise rate, although there is a finite density of flux excitations in the steady state.

\begin{figure}
\begin{centering}
\includegraphics[width=(\textwidth-\columnsep)/2]{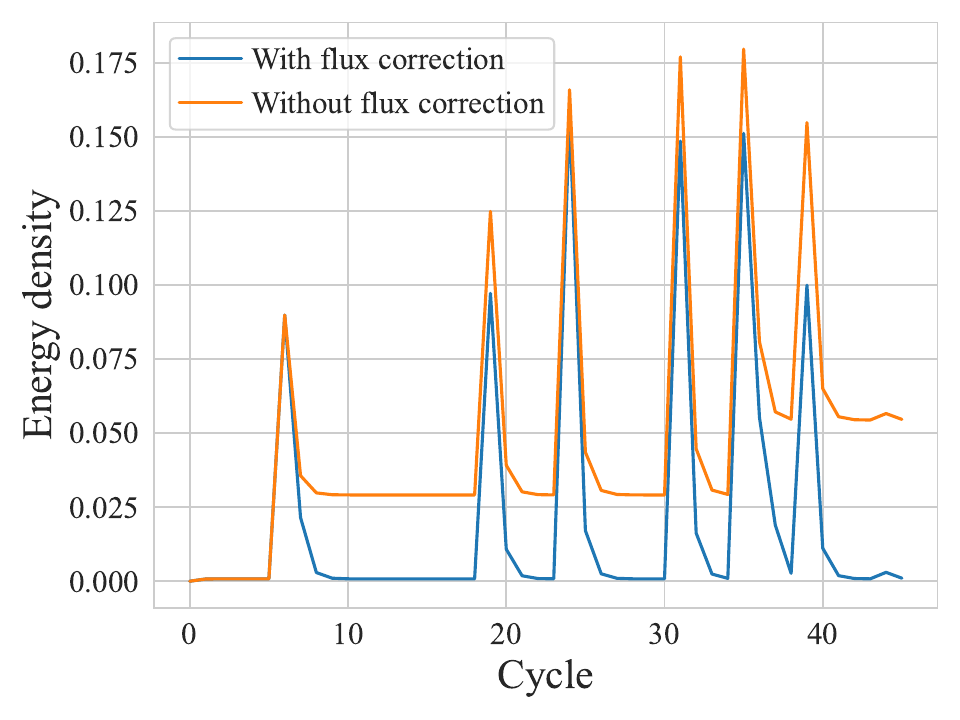}
\par\end{centering}
\caption{\textbf{Cooling in the presence of noise.} A KSL system of size $6\times 6$ unit cells initialized in its ground state is cooled in the presence of a decoherence rate of $\zeta=10^{-3}$ errors per cycle per qubit. The cycle duration is set to $T=20$ and $N_t=800$ Trotter steps are used per cycle. The energy density is plotted at the end of each cycle. The case where fluxes are corrected every cycle is shown in blue and the case where fluxes are left uncorrected is in orange.}
\label{fig. single time trace}
\end{figure}

We have simulated a KSL system of size $6\times 6$ unit cells, cooled in the presence of decoherence for multiple cycles. The results are presented in Fig.~\ref{fig. single time trace}. The energy density is shown at the end of each cycle, with and without the active correction of the flux degrees of freedom.  
In the simulation with flux correction, the fluxes are measured and removed at the end of each cycle. For simplicity, we have assumed that the measurements are ideal, and the flux correction operations are perfect.
In both cases the stochastic action of errors appear as local peaks in the energy density after which the fermionic excitations produced are cooled. However, when errors which excite the flux degrees of freedom occur and these are not corrected, further cycles of fermionic cooling are only able to reach the ground state within the new excited flux sector. After many cycles, a steady state is reached in which the flux degrees of freedom are in a maximally mixed state.

\begin{figure}
\begin{centering}
\includegraphics[width=(\textwidth-\columnsep)/2]{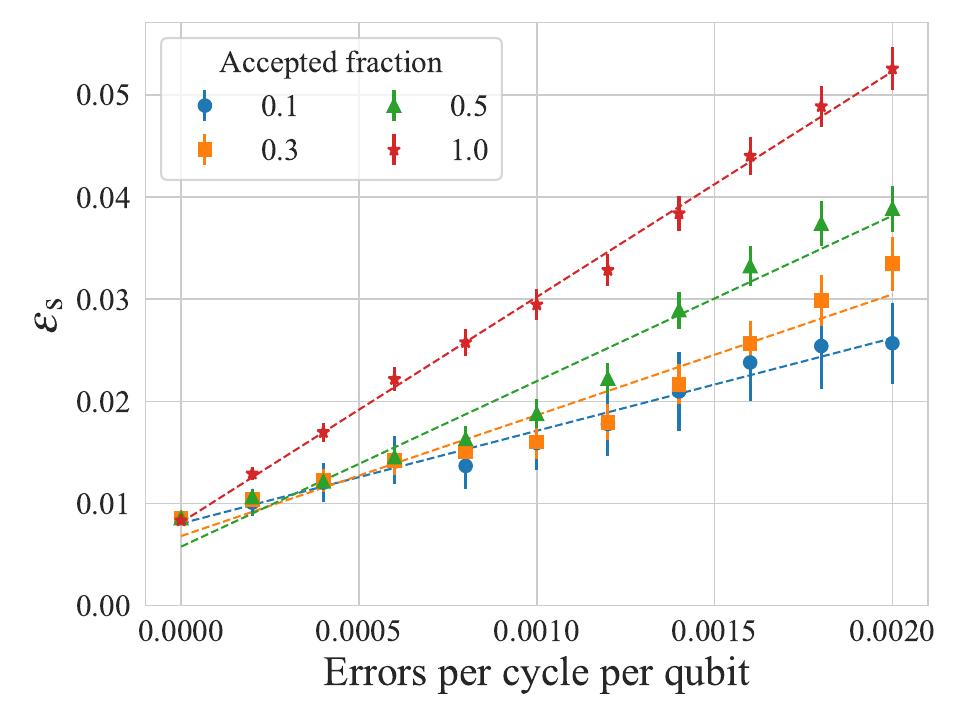}
\par\end{centering}
\caption{\textbf{Steady state in the presence of noise.} The energy density of a KSL model of size $10\times 10$ unit cells in the steady state reached after many cooling cycles as a function of the rate of depolarizing noise. The energies are averaged over $2000$ cycles per data point and error bars represent the standard error of the mean determined for an effective sample size of $2000/(2 n_0)$ taking into account exponentially decaying correlations over approximately $n_0=3$ cycles. The cycle duration is set to $T=10$ and $N_t=400$ Trotter steps are used per cycle. The flux degrees of freedom are monitored and corrected at the end of each cycle. Post selection is performed with respect to the weight of decoded Pauli correction applied to fix the detected flux excitations, exponentially averaged over recent cycles $\bar{w}_n$. Cycles requiring Pauli corrections of high weight are omitted. Different colors correspond to different fractions of cycles accepted as ranked by this criterion. The data is fit by linear functions (dashed lines).}
\label{fig. vs. error rate}
\end{figure}

Figure \ref{fig. vs. error rate} shows the energy density of the steady state reached by cooling a system of size $10\times 10$ in the presence of noise while correcting the flux degrees of freedom at each cycle. The energy density is linear in the error rate, demonstrating that cooling of both fermionic and flux excitations is done at a rate proportional to their density.

The active correction of the flux degrees of freedom by measurement and feedback provides a natural tool for further error mitigation by post selection. In cycles which require a large weight Pauli correction in order to annihilate the monitored flux excitations, the error together with its correction likely result in the insertion of fermionic excitations which must be removed by further cooling cycles. We denote the weight of the Pauli correction (defined as the number of gates applied to correct the fluxes) at cycle $n$ by $w_n$, and calculate its exponential moving average
\begin{align}
\label{eq: ema of pauli correction}
\bar{w}_n = \frac{\sum_{m\leq n} w_m e^{-\frac{n-m}{n_0}}}{\sum_{m\leq n} e^{-\frac{n-m}{n_0}}},
\end{align}
where $1/n_0$ is the exponential cooling rate in the absence of noise. We perform post selection in Fig.~\ref{fig. vs. error rate} by accepting a fraction of the cycles with the smallest values of $\bar{w}_n$, using $n_0=3$.
Performing such post selection succeeds in reducing the expectation value of the energy density. In the limit of a very low accepted fraction of the data, the energy density is reduced approximately by a factor of two at low error rates. In this limit, all cycles in which $\sigma$ spins are affected by errors are removed, but cycles in which $\tau$ spins are disturbed remain part of the data. Further post selection can be done by imposing a requirement on the outcomes of the measurements of the $\tau$ degrees of freedom from cycle to cycle \cite{mathhies2022adibatic_demag}.

\subsection{Extension to arbitrary fermionic Hamiltonians}
\label{sec: arbitrary fermions}

Our protocol for preparing the ground state of the KSL model can be generalized as a method for preparing the ground state of an arbitrary (interacting) fermionic Hamiltonian in 2D. Given some local target Hamiltonian acting on the system Majorana fermions $c_\mathbf{J}^z$, we introduce the gauge fields $u_{\mathbf{I},\mathbf{J}}$ artificially and modify the target Hamiltonian such that the $c_\mathbf{J}^z$ fermions become charged under the gauge field. 
Namely, we modify each hopping term between site $\mathbf{I}$ and site $\mathbf{J}$ in the target Hamiltonian by the product of the gauge fields on a path connecting the two sites. 
We then introduce the bath fermions $c_\mathbf{J}^x,c_\mathbf{J}^y$, 
and use Eq.~(\ref{eq: mapping spins to fermions}) to map the fermionic Hamiltonian (including the hopping between the system and bath fermions, used for cooling the system) onto a bosonic local Hamiltonian acting on the $\sigma$ and $\tau$ spins. 
The ground state of the Hamiltonian is prepared according to the protocol above and this state is related to the ground state of the target Hamiltonian by the known gauge transformation as discussed in Sec.~``Preparing a chiral spin liquid''.

Generalizations to other problems, e.g., with complex spinful fermions or multiple orbitals per site, are straightforward. We also note that Hamiltonians with fermion number conservation can be implemented within our protocol; the fermion number changes during each cycle, but the system is cooled towards the ground state that has a well-defined fermion number set by a chemical potential term. 

\section{Discussion}

In this work, we have presented an efficient protocol for preparing certain chiral topological states of matter on quantum simulators. Our prime example is the chiral phase of the Kitaev spin model on the honeycomb lattice. Our protocol can be similarly used to prepare invertible topological phases of fermions, such as Chern insulators and chiral topological superconductors. Importantly, our protocol does not require an a-priori knowledge of the ground state wavefunction, and should perform equally well in the presence of interactions (although the phases we prepare have a description in terms of non-interacting fermions). In the absence of interactions and decoherence, the ground state can be reached after a single cycle, up to adiabatic errors 
(for the non-interacting case, a similar approach was proposed in Ref.~\cite{tantivasadakarn2022hierarchy}). 


Our scheme utilizes a simultaneous fermionization of the target system and the cooling bath with which it is coupled, in order to allow the removal of single fermionic excitations from the system to the bath. Crucially, the protocol relies on repeated measurements of the bath spins that detects the fermionic excitations, and local classical feedback that removes them. 

 In practice, for running this algorithm on real quantum hardware, one should choose the time duration $T$ of the unitary evolution and the number of Trotter steps, $N_t$, to optimize the trade-off between diabatic errors and Trotterization noise (which decrease upon increasing $T$ and $N_t$) versus the noise deriving from environmental decoherence and low fidelity gates (which are typically proportional to the number of gates applied, and hence to $N_t$). The optimization can be done even without knowledge of the noise characteristics, by minimizing the measured energy of the steady state with respect to the protocol parameters, such as $T$ and $N_t$. 

Our protocol offers a parametric advantage compared to other recently-proposed methods to prepare chiral 2D states. Ref. \cite{Kalinowski_2023} describes a method to prepare the chiral state of the Kitaev honeycomb model by adiabatic evolution. Due to the gap closing along the path, the excitation energy of the final state (or the ground state fidelity) scales polynomially with the duration of the time evolution. In contrast, in our protocol, no adiabatic path is required, and the excitation energy scales exponentially with the cycle duration. In Ref.~\cite{chen2024sequential}, a method for preparing chiral states using sequential quantum circuits is described; this method requires a circuit whose depth scales with the linear size of the system in order to reach a state with a given energy density. 
Ref.~\cite{tantivasadakarn2022hierarchy} proposes to prepare a ``quantum double'' state composed of the chiral state of interest and its time-reversed partner. The doubled state supports a gapped boundary, and can be prepared using measurements and feedback in one shot. Then, a finite-depth quasi-local unitary evolution can decouple the state from its time reversal partner. This idea has some similarities to the method we propose both in terms of the use of non-unitary operations and the scaling achieved. However, the routes are distinct and, in particular, here we give an explicit detailed construction of the protocol whereas in Ref.~\cite{tantivasadakarn2022hierarchy} the idea is presented and justified on general theoretical grounds without providing an explicit protocol for the unitary evolution step (as would be needed to enable an implementation in practice). 


We emphasize that, while our method does not require knowing the ground state in advance, the performance of the protocol depends crucially on the topological phase to which the ground state belongs. Developing practical and efficient methods to prepare ground states in more complicated topological phases, such as fractional Chern insulators, 
is an interesting outstanding challenge. 


\section{Methods}

\subsection{Removing fluxes in 2D}
\label{app: fluxes}
In order to reach the ground state of the 2D KSL, one should prepare the system in a flux-free state of the gauge field $u$ and then proceed to cool the fermionic excitations $c^z$. 
Fluxes can be removed by measuring the plaquette operators $W_{i,j}$ which correspond to local 6-body spin operators shown in Eq.~\eqref{eq: fluxes in terms of spins}. 
The excitations $W_{i,j} = -1$ can be removed in pairs by products of local unitary operations. We note that the operator $\sigma^\alpha_{\mathbf{J}}$ at a vertex of the honeycomb lattice shared between three hexagons anticommutes with two of the three flux operators on these hexagons,
while commuting with the third (as well as all the other plaquettes in the lattice). Therefore, applying such a unitary operator $\sigma^\alpha_{\mathbf{J}}$ flips the signs of the two corresponding plaquettes, potentially also inserting fermionic excitations. Choosing some pairing of the measured excited plaquettes, one can construct a string of spin operators $\prod_{\mathbf{J}\in S}\sigma^{\alpha(\mathbf{J})}_{\mathbf{J}}$ connecting 
each pair such that the product anti-commutes only with the 
plaquette operators at the end points of the string.

Using the above approach, a finite density of flux exitations can be corrected using measurements and feed-forward unitaries (analogously to the error correction protocol in the surface code~\cite{Dennis_2002}). The composite error (the original error affecting the fluxes together with the correction) is a product of Pauli operators acting on the $\sigma$ spins which commutes with all of the flux operators, leaving them unaffected. However, the action of this operator on the system excites a number of residual fermionic excitations proportional to its weight (the number of spins on which it acts). Therefore, assuming a low error rate resulting in a low weight Pauli error with high probability, it is important to construct a correction with a low weight as well by annihilating the flux excitations in local pairs - using a minimum weight perfect matching decoder \cite{edmonds_1965}. Otherwise, the density of residual fermionic excitations will scale with system size.

The resulting fermionic excitations left after fusing the fluxes can be removed later by the coupling to the bath. Consequently, in the presence of noise, the combined algorithm (composed of a unitary sweep, measurement of the bath spins and the fluxes, and feed-forward correction of the fluxes) should reach a steady state whose energy density is proportional to the noise rate and is independent of system size. This is the algorithm used in presence of noise, whose results are shown in Figs.~\ref{fig. single time trace} and \ref{fig. vs. error rate}.

\subsection{Smooth evolution of time dependent couplings}
\label{app: smooth curves}

In order to approach the adiabatic limit exponentially with increasing cooling cycle duration $T$, we choose the time dependence of the couplings $B$ and $g$ to have finite derivatives at all orders. Explicitly, we use a smooth step function $\mathcal{S}(x)$ as used in Ref.~\cite{kishony2023gauged}, defined as
\begin{align}
\mathcal{S}(x) = \begin{cases}
			0, & x<0\\
            1, & x>1\\
            \frac{1}{1+e^{\frac{1}{x}+\frac{1}{x-1}}}, & \text{otherwise}
		 \end{cases}.
\end{align}
Using this, we express $B(t)$ and $g(t)$ as
\begin{align}
B(t)=B_0\left[1-\mathcal{S}\left(\frac{t}{T}\right)\right],
\end{align}
\begin{align}
g(t)=g_1\mathcal{S}\left(\frac{t}{t_1}\right)\left[1-\mathcal{S}\left(\frac{t-t_2}{T-t_2}\right)\right].
\end{align}
These curves are shown in Fig.~\ref{fig. g and B} 
for $t_1=\frac{1}{4}T$, $t_2=\frac{3}{4}T$ as chosen in our numerical simulations.

\subsection{Single-particle density matrix simulation of free fermions}
\label{app: free fermion single-particle density matrix}

The free evolution of a quantum many-body system under a quadratic fermionic Hamiltonian, including measurements and resets of single sites and Pauli errors, can be traced numerically efficiently, requiring only 
time and memory that scale as polynomials of the system size. Here, we briefly describe the technique used to simulate the noisy dynamics of the cooling process discussed in Sec.~``Performance with noise''. 
The technique is the same as that used in Ref. \cite{kishony2023gauged}.

We keep track of the single particle density matrix $R_{\mathbf{I},\mathbf{J}}^{\alpha,\beta}=\left\langle ic_{\mathbf{I}}^{\alpha}c_{\mathbf{J}}^{\beta}\right\rangle $ as well as the values of the $\mathbb{Z}_2$ gauge field $u_{\mathbf{I},\mathbf{J}}=\pm 1$ as they evolve under the noisy cooling dynamics.
This allows us to calculate any quadratic observables, and the energy in particular.



During free evolution under the Hamiltonian Eq.~(\ref{eq:H_KSL_fermions})
of the main text within a fixed sector of the gauge field $u$, the single particle density matrix evolves under a unitary matrix $W(t)$ by conjugation, $R(t)=W^{\dagger}(t)R(0)W(t)$. This unitary is given by the time ordered exponentiation of the Hamiltonian in matrix form, and can be found numerically by Trotterization or using an ordinary differential equation solver. In Sec.~``Performance with noise'' of the main text 
we use second order Trotterization, making $N_t$ steps per cooling cycle. Importantly, the unitary depends on the values of the gauge field which are looked up for its computation.

The resetting operation of the bath fermions at the end of each cycle $ic_\mathbf{J}^yc_\mathbf{J}^x\rightarrow 1$ is handled as in Ref.~\cite{kishony2023gauged}. This amounts to setting all matrix elements of $R_{\mathbf{I},\mathbf{J}}^{\alpha,\beta}$ in the two rows and the two columns corresponding to $\mathbf{J},x$ and to $\mathbf{J},y$ to zero, and then setting the $2\times2$ block at their intersection to the values found at $ic_\mathbf{J}^yc_\mathbf{J}^x=1$.
Depolarizing noise is implemented through stochastic application of Pauli errors before each Trotter step of the time evolution and for each spin ($\{\sigma_\mathbf{J},\tau_\mathbf{J}\}$) with probability of $\zeta/N_t$. The Pauli $\tau_\mathbf{J}^\alpha$ operators are bilinear in the $c$ fermions [see Eq.~\eqref{eq: mapping spins to fermions}], such that they correspond to evolution under a quadratic fermionic Hamiltonian. Pauli errors $\sigma_\mathbf{J}^\alpha$ act on the $b$ Majorana fermions, flipping the signs of two of the three gauge fields incident to site $\mathbf{J}$. 

Measuring the flux degrees of freedom within this framework simply requires reading off the product of the values of the $\mathbb{Z}_2$ gauge fields around the corresponding plaquettes. These values are well defined at every point in time for each trajectory in these simulations, and their fluctuations can be obtained by averaging over multiple trajectories with stochastic noise. Correcting these fluxes in order to return to the flux-free sector, as described in Methods subsection ``Removing fluxes in 2D'', 
is done by applying a unitary correction given by a product of Pauli $\sigma_\mathbf{J}^\alpha$ operators. 




\acknowledgements

G.K. and E.B. were supported by CRC 183 of the Deutsche Forschungsgemeinschaft (subproject A01), a research grant from
the Estate of Gerald Alexander, and ISF-MAFAT Quantum Science and Technology Grant no. 2478/24. 
 M.R. acknowledges the Brown Investigator Award, a program of
the Brown Science Foundation, the University of Washington College of Arts and Sciences, and the Kenneth K.
Young Memorial Professorship for support.

\section{Author Contributions}
G.K., M.S.R., and E.B.~designed the research and contributed to writing the paper. G.K.~performed the analytical and numerical calculations.

\section{Competing interests}
The authors declare no competing interests.

\section{Data availability}
The data that support the findings of this study are available at \url{https://github.com/kishonyWIS/free_fermions}.

\section{Code availability}
The code used in this work is available at \url{https://github.com/kishonyWIS/free_fermions}.


\bibliography{References}

\end{document}


\title{Supplementary Material for: \\ 
Efficiently preparing chiral states via fermionic cooling on bosonic quantum hardware}

\author{Gilad Kishony}\email{gilad.kishony@weizmann.ac.il}
\affiliation{Department of Condensed Matter Physics,
Weizmann Institute of Science,
Rehovot 76100, Israel}
\author{Mark S.~Rudner}
\affiliation{Department of Physics, University of Washington, Seattle, WA 98195-1560, USA}
\author{Erez Berg}
\affiliation{Department of Condensed Matter Physics,
Weizmann Institute of Science,
Rehovot 76100, Israel}

\maketitle

\section*{Supplementary Note 1: Analysis in momentum space}
\label{app: momentum space}

In this section we study the cooling dynamics of the quadratic fermionic Hamiltonian in Eq.~(8) in the main text 
in the flux-free sector with uniform $\tau^z_{\mathbf{J}}=1$. We choose a gauge where $u_{\mathbf{I},\mathbf{J}}=1$, for which the Hamiltonian is translationally invariant,  and Fourier transform the fermionic Hamiltonian to find
\begin{align}
&H=\sum_{\bm{k}}\bm{\Psi}_{\bm{k}}^\dagger H_{\bm{k}}\bm{\Psi}_{\bm{k}},
\end{align}
where
$\bm{\Psi}_{\bm{k}}^\dagger=\begin{pmatrix}c_{(\bm{k},A)}^{z\dagger}&
c_{(\bm{k},B)}^{z\dagger}&
c_{(\bm{k},A)}^{y\dagger}&
c_{(\bm{k},B)}^{y\dagger}&
c_{(\bm{k},A)}^{x\dagger}&
c_{(\bm{k},B)}^{x\dagger}
\end{pmatrix}$,
with the $6 \times 6$ time-dependent Hamiltonian describing the modes at each $\bm{k}$ given by 
\begin{align}
    H_{\bm{k}}(t)=\left(\begin{array}{ccc}
h_{\bm{k}} & G(t) & 0\\
G^{\dagger}(t) & 0 & -iB(t)\mathbb{I}\\
0 & iB(t)\mathbb{I} & 0
\end{array}\right).
\label{eq:Hk}
\end{align}
Here, $\mathbb{I}$ is a $2\times 2$ unit matrix, 
$h_{\bm{k}}$ is the $2 \times 2$ submatrix describing the ``system'' ($c^z_{(\bm{k},s)}$) fermions, 
\begin{align}
h_{\bm{k}}&=\begin{pmatrix} \Delta(\bm{k}) & if(\bm{k})\\
-if(-\bm{k}) & -\Delta(\bm{k})
\end{pmatrix},
\label{eq:h_k}
\end{align}
and the $2 \times 2$ matrix $G$ that describes hopping between fermionic system and bath ($c^y_{(\bm{k},s)}$) is given by
\begin{align}
G(t)=&\begin{pmatrix} -ig(t)\Gamma_A & 0\\
0 & -ig(t)\Gamma_B
\end{pmatrix},
\end{align}
where we allow the coupling $g_{\J}$ to be sublattice dependent $g_{(i,j,s)}(t) = g(t-t_n)\Gamma_s$ rather than spatially uniform as chosen in the main text. 
The functions $\Delta(\bm{k})$ and $f(\bm{k})$ in Supplementary Eq.~(\ref{eq:h_k}) are given by
$\Delta(\bm{k})=2\kappa\left[\sin{k_x}-\sin{k_y}+\sin{(k_y-k_x)}\right]$
and $f(\bm{k})~=~\mathcal{J}_{x}+\mathcal{J}_{y}e^{-ik_{x}}+\mathcal{J}_{y}e^{-ik_{y}}$.

\subsection{Adiabatic limit}
\label{sec: abiabatic}

For the cooling protocol to succeed, it is essential to ensure that for all $\bm{k}$ the off-diagonal terms $G$ have a non-zero matrix element between the ground state of the target Hamiltonian, $h_{\bm{k}}$, and the ground state of the Hamiltonian $B\sigma^y$ of one of the bath modes, $c^{x}_{(\bm{k},s)},c^{y}_{(\bm{k},s)}$. Importantly, if the system-bath coupling 
is nonvanishing for all $\bm{k}$, 
in the perfectly adiabatic limit  the system reaches its ground state after a single cycle (see discussion in Ref.~\cite{kishony2023gauged}). 
The instantaneous spectrum of $H_{\bm{k}}$ in the topological phase with $\mathcal{J}_x=\mathcal{J}_y=\mathcal{J}_z$ and $\kappa>0$ (where the ground state of the system fermions  $c^z_{(\bm{k},s)}$ carries a Chern number of $1$ when $g=0$) is shown at the point $\bm{k}=(2\pi/3,-2\pi/3)$ as a function of $B$ in Supplementary Fig.~\ref{fig. KSL spectrum}. At this point, we find $f(\bm{k})=0$ and $h_{\bm{k}}\propto\sigma^z$, such that
using only the $A$ sublattice bath (taking $\Gamma_B=0$) does not allow cooling of fermionic excitations at this $\bm{k}$ point.

\begin{figure}
\begin{centering}
\includegraphics[width=(\textwidth-\columnsep)/2]{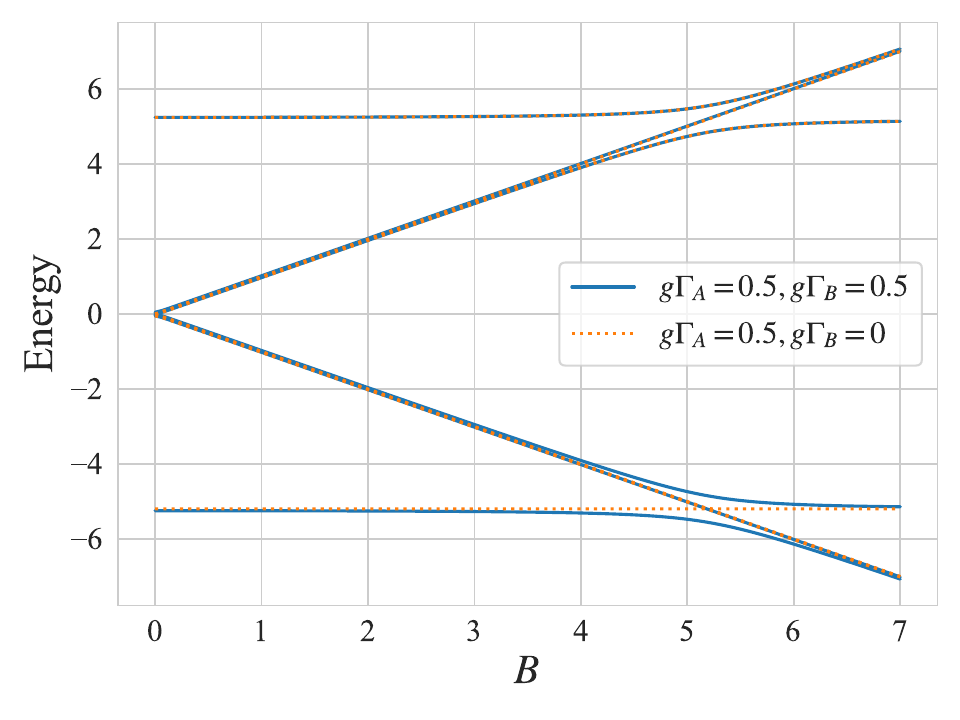}
\par\end{centering}
\caption{\textbf{The instantaneous single particle excitation spectrum of the cooled KSL.} The spectrum is drawn for the model in the topological phase with $\mathcal{J}=1,\kappa=1$ at $\bm{k}=(2\pi/3, -2\pi/3)$ as a function $B$. Solid lines correspond to $g\Gamma_A=g\Gamma_B=0.5$, while for the dashed lines $g\Gamma_A=0.5$ and $g\Gamma_B=0$ such that only the $A$ sublattice bath is cooling the system. In the case of two cooling sublattices, one of the two degenerate modes of the bath is mapped adibatically to the eigenstate of the system (the lowest solid curve) such that the system reaches its ground state at the end of the evolution. With a single cooling site per unit cell, the bath remains decoupled from the system at this $\bm{k}$ point.}
\label{fig. KSL spectrum}
\end{figure}

The existence of this gap closing is due to the non-zero Chern number of the target ground state. As $\bm{k}$ varies, the spinor corresponding to the system's ground state covers the entire Bloch sphere. Therefore, there must be a value of $\bm{k}$ for which the system's ground state is orthogonal to the spinor formed by acting with the first column of $G$ on the bath's ground states. At this $\bm{k}$ point, the coupling to 
the bath on the $A$ sublattice alone cannot cool the system. 

Using two bath sublattices is enough to overcome this problem. As long as matrix $G$ is not singular, the spinors obtained by acting with the columns of $G$ on the system's ground state are not identical, and at least one of these spinors has a non-zero overlap with the system's ground state at each $\bm{k}$. Note that, in contrast to the topological phase, in the trivial phase where the Chern number vanishes (obtained, e.g., when $\mathcal{J}_z >> \mathcal{J}_x, \mathcal{J}_y$ and $\kappa=0$), one bath sublattice would generically be sufficient (with appropriately chosen coupling). 

\subsection{Diabatic corrections}
\label{sec: diabatic}

Deviations from the adiabatic limit have two main effects. First, the protocol
does not converge to its steady state after a single cycle, but rather
converges to the steady state exponentially with the number of cycles
performed. 
Second, the steady state deviates from the ground state of the
system. Importantly, we shall now show that the deviation of the steady
state from the ground state due to diabatic errors is controlled by
the intrinsic energy gap of the system, and \emph{not} by the avoided crossing
in the middle of the cycle (see Supplementary Fig. \ref{fig. 0d tfi spectrum vs time}). The mid-cycle avoided crossing gap controls
the rate of convergence to the steady state.

Returning to the case of $\Gamma_A=\Gamma_B=1$, the diabatic corrections are most straightforward to analyze after
a series of unitary transformations which map the problem at each $\bm{k}$ into an evolution of three spins.
First, we perform a $\bm{k}-$dependent unitary transformation 
\begin{align}
U_{\bm{k}} = \begin{pmatrix}V_{\bm{k}}&&\\
&V_{\bm{k}}&\\
&&V_{\bm{k}}\\
\end{pmatrix},
\end{align}
where $V_{\bm{k}}$ is chosen such that $V_{\bm{k}} h_{\bm{k}} V_{\bm{k}}^\dagger = E_{\bm{k}}\sigma^y$,
with $E_{\bm{k}}=\sqrt{\left|\Delta(\bm{k})\right|^2+\left|f(\bm{k})\right|^2}$ being the spectrum of the KSL model. 
This brings the Hamiltonian
[Supplementary Eq.~(\ref{eq:Hk})] into a purely antisymmetric form
\begin{align}    \widetilde{H}_{\bm{k}}=U_{\bm{k}}H_{\bm{k}}U_{\bm{k}}^{\dagger}=\left(\begin{array}{ccc}
    E_{\bm{k}}\sigma^{y} & -ig(t)\mathbb{I} & 0\\
    ig(t)\mathbb{I} & 0 & -iB(t)\mathbb{I}\\
    0 & iB(t)\mathbb{I} & 0
    \end{array}\right).
    \label{eq:Hk_tilde}
\end{align}

\begin{figure}
\begin{centering}
\includegraphics[trim={0.4cm 0.7cm 1.3cm 1.5cm},clip,width=(\textwidth-\columnsep)/2]{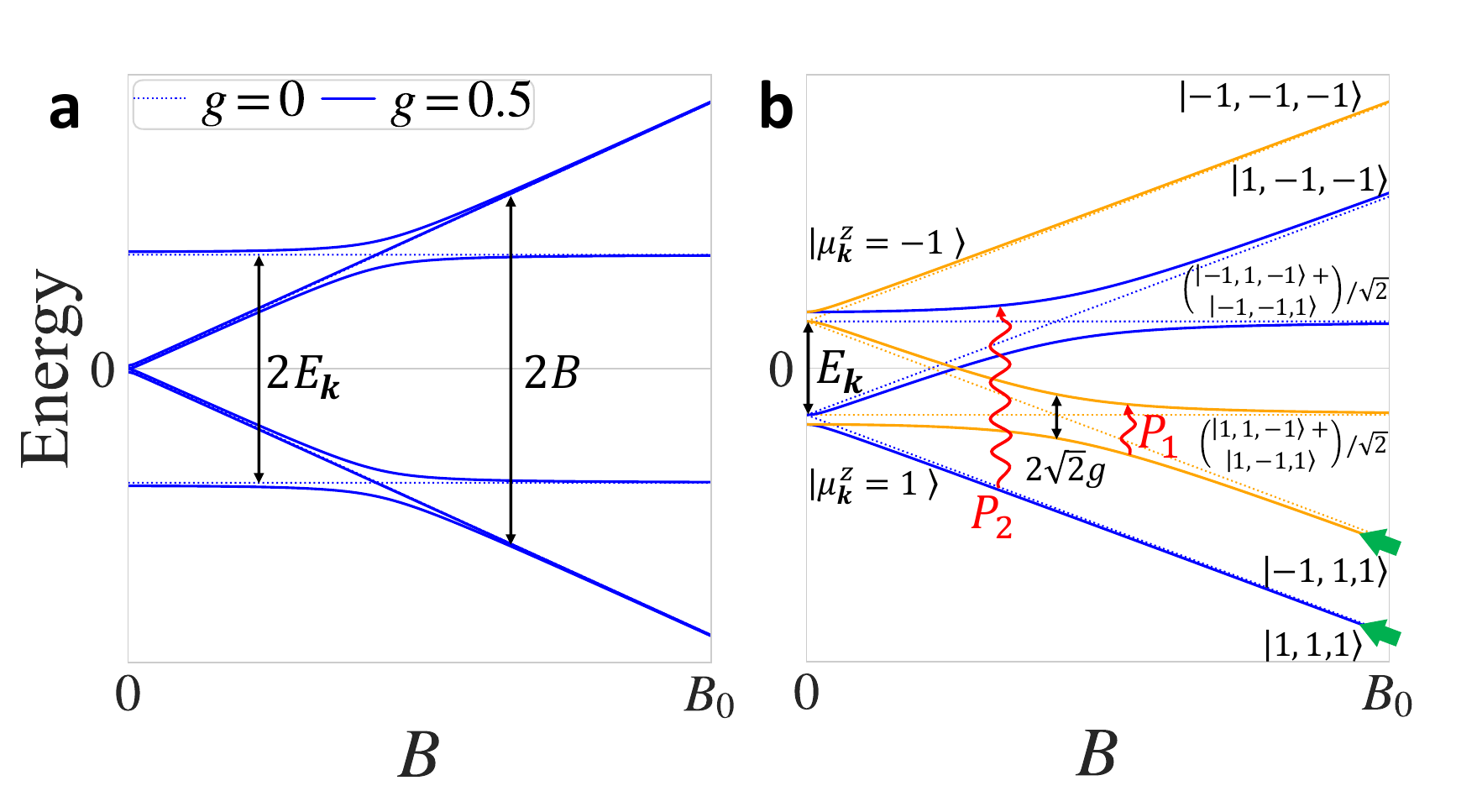}
\par\end{centering}
\caption{\textbf{The instantaneous spectrum of the 0D model at each $\bm{k}$ point during the adiabatic sweep of the cooling protocol.} Dotted lines show the same spectrum when the coupling $g$ is held at $0$. \textbf{(a)} 
Single particle excitation spectrum of the Hamiltonian $H_{\bm{k}}(t)$ in Supplementary Eq.~\eqref{eq:Hk}. \textbf{(b)} 
Spectrum of the simplified spin model described by Supplementary Eq.~\eqref{eq: 0D model Hamiltonian}, colored blue and orange 
according to the parity of the states; decoupled states are omitted. 
The adiabatic sweep is performed after the reset (green arrows) which initializes the state in a superposition of $\left|\mu_{\bm{k}}^{z},\eta_{\bm{k}}^{z},\chi_{\bm{k}}^{z}\right\rangle=\left|1,1,1\right\rangle$ and $\left|-1,1,1\right\rangle$ at time $t=0$. In the adiabatic limit, the states $\left|-1,1,1\right\rangle$ and $\frac{1}{\sqrt{2}}\left(\vert 1,1,-1\rangle+\vert 1,-1,1\rangle\right)$ are exchanged at a Landau-Zener transition, such that the ground state $\mu_{\bm{k}}^z=1$ is reached. 
Diabatic transitions with associated probabilities $P_{1,\bm{k}}$ and $P_{2,\bm{k}}$ are illustrated in red.}
\label{fig. 0d tfi spectrum vs time}
\end{figure}

Next, we write the corresponding vector of complex fermionic annihilation operators
$\widetilde{\bm{\Psi}}_{\bm{k}}=U_{\bm{k}}\bm{\Psi}_{\bm{k}}$ in terms of two vectors of Majorana
operators: $\widetilde{\bm{\Psi}}_{\bm{k}}=\frac{1}{2}(\bm{a}_{\bm{k}}+i\bm{d}_{\bm{k}})$, where $\bm{a}_{\bm{k}}^{T}=(a_{1,\bm{k}},\dots,a_{6,\bm{k}})$
with $a_{n,\bm{k}}=a_{n,\bm{k}}^{\dagger}$, and similarly for $\bm{d}_{\bm{k}}$.
Using the antisymmetric property of $\widetilde{H}_{\bm{k}}$, the many-body
Hamiltonian decouples when written in terms of $\bm{a}_{\bm{k}}$, $\bm{d}_{\bm{k}}$:
\begin{equation}
H=\frac{1}{4}\sum_{\bm{k}}\left[\bm{a}_{\bm{k}}^{T}\widetilde{H}_{\bm{k}}\bm{a}_{\bm{k}}+\bm{d}_{\bm{k}}^{T}\widetilde{H}_{\bm{k}}\bm{d}_{\bm{k}}\right].
\end{equation}
For a given $\bm{k}$, 
the parts of the Hamiltonian that
depend on $\bm{a}_{\bm{k}}$ and $\bm{d}_{\bm{k}}$ are independent and identical, and it is sufficient to analyze one of them. 


\subsubsection{Mapping to a three spin problem}
\label{sec: mapping to three spin}

The part that
depends on $\bm{a}_{\bm{k}}$ 
can be written in terms of three pseudospin operators, 
defined
such that $\mu_{\bm{k}}^{z}=ia_{1,\bm{k}}a_{2,\bm{k}}$, $\eta_{\bm{k}}^{z}=ia_{3,\bm{k}}a_{5,\bm{k}}$, $\chi_{\bm{k}}^{z}=ia_{4,\bm{k}}a_{6,\bm{k}}$. 
The basis of the corresponding Hilbert space can be labelled according to the eigenvalues of $\mu_{\bm{k}}^{z},\eta_{\bm{k}}^z,\chi_{\bm{k}}^z$; we denote the basis states as $\vert \mu_{\bm{k}}^{z},\eta_{\bm{k}}^z,\chi_{\bm{k}}^z \rangle$. 
The $\bm{a}_{\bm{k}}$-dependent part of the
Hamiltonian is written as 
\begin{equation}
H_{\bm{k}}^{a}(t)=-\frac{1}{2}E_{\bm{k}}\mu_{\bm{k}}^{z}-\frac{1}{2}B(t)\left(\eta_{\bm{k}}^{z}+\chi_{\bm{k}}^{z}\right)+\frac{1}{2}g(t)\mu_{\bm{k}}^{x}\left(\eta_{\bm{k}}^{x}+\chi_{\bm{k}}^{x}\right).
\label{eq: 0D model Hamiltonian}
\end{equation}
We have thus mapped the problem into that of three coupled spins. The Hamiltonian of the $\bm{d}_{\bm{k}}$ sector has
the same form, and can be treated similarly. 
Note that the $\mu_{\bm{k}},\eta_{\bm{k}},\chi_{\bm{k}}$ spins are distinct from the physical ones, $\sigma_{\bm{J}}$ and $\tau_{\bm{J}}$.

\subsubsection{Solution of the three spin problem}
\label{sec: solution three spin problem}

At $g=0$ the Hamiltonian is diagonalized in the orthonormal basis $\frac{1}{\sqrt{2}}\left(\vert\pm 1,1,-1\rangle-\vert\pm 1,-1,1\rangle\right)$, $\vert\pm 1,1,1\rangle$, $\frac{1}{\sqrt{2}}\left(\vert\pm 1,1,-1\rangle+\vert\pm 1,-1,1\rangle\right)$, $\vert\pm 1,-1,-1\rangle$ with energies given by $\mp E_{\bm{k}}/2$, $\mp E_{\bm{k}}/2-B$, $\mp E_{\bm{k}}/2$, $\mp E_{\bm{k}}/2+B$. At finite $g$, the first two states remain decoupled. The Hamiltonian in the complementary subspace written in this basis is given by

\begin{align}
H=\begin{pmatrix}
-\frac{E_{\bm{k}}}{2}-B &  &  & \sqrt{2}g & &\\
 & \frac{E_{\bm{k}}}{2}-B & \sqrt{2}g & & & \\
& \sqrt{2}g & -\frac{E_{\bm{k}}}{2} & & & \sqrt{2}g\\
\sqrt{2}g & & & \frac{E_{\bm{k}}}{2} & \sqrt{2}g & \\
& & & \sqrt{2}g & -\frac{E_{\bm{k}}}{2}+B & \\
& & \sqrt{2}g & & & \frac{E_{\bm{k}}}{2}+B \\
\end{pmatrix},
\end{align}
and its spectrum is shown as a function of $B$ for $g>0$ (solid lines) and for $g=0$ (dotted lines) in Supplementary Fig.~\ref{fig. 0d tfi spectrum vs time}b. Turning on a non-zero $g$ couples $\frac{1}{\sqrt{2}}\left(\vert\pm 1,1,-1\rangle+\vert\pm 1,-1,1\rangle\right)$ to $\vert\mp 1,1,1\rangle$ and $\vert\mp 1,-1,-1\rangle$. 
(Note that $[\mu_{\bm{k}}^z \eta_{\bm{k}}^z \chi_{\bm{k}}^z, H^a_{\bm{k}}]=0$: the even and odd parity sectors of total $z$ pseudospin are decoupled.) 


Each cooling cycle is initialized with the bath spins $\eta_{\bm{k}}$, $\chi_{\bm{k}}$ in their respective ground states, i.e., the system is initialized in some mixture of the states $\vert\pm 1,1,1\rangle$. For $g=0$, the energies of the eigenstates $\left|-1,1,1\right\rangle $ and $\frac{1}{\sqrt{2}}\left(\vert 1,1,-1\rangle+\vert 1,-1,1\rangle\right)$ cross at $B=E_{\bm{k}}$, while the state $\left|1,1,1\right\rangle$ remains spectrally separated from the only state it is directly coupled to, $\frac{1}{\sqrt{2}}\left(\vert -1,1,-1\rangle+\vert -1,-1,1\rangle\right)$. With a finite $g$, the crossing becomes an avoided crossing and in the adiabatic limit, the states $\vert-1,1,1\rangle$ and $\frac{1}{\sqrt{2}}\left(\vert 1,1,-1\rangle+\vert 1,-1,1\rangle\right)$ are exchanged under the time evolution, while $\left| 1,1,1 \right\rangle$ 
evolves trivially.

Away from the adiabatic limit, we denote the diabatic transition probability for the $\left| -1, 1, 1\right\rangle$, $\frac{1}{\sqrt{2}}\left(\vert 1,1,-1\rangle+\vert 1,-1,1\rangle\right)$ pair as $P_{1,\bm{k}}$ and the diabatic transition probability 
between $\left| 1,1,1\right\rangle$ and $\frac{1}{\sqrt{2}}\left(\vert -1,1,-1\rangle+\vert -1,-1,1\rangle\right)$ as $P_{2,\bm{k}}$ (see Supplementary Fig.~\ref{fig. 0d tfi spectrum vs time}b).
For $g(t)/E_{\bm{k}} \ll 1$ and assuming $g(t)$ is constant for times where $B(t) \approx E_{\bm{k}}$, $P_{1,\bm{k}}$ is approximately given by the Landau-Zener formula, $P_{1,\bm{k}}\approx e^{-c_1 Tg^2/B_0}$ for some constant $c_1$. The second transition probability, 
$P_{2,\bm{k}}$, is determined by the single particle gap of the target Hamiltonian, $P_{2,\bm{k}}\approx e^{-c_2 TE_{\bm{k}}^2/B_0} \ge e^{-c_2 TE_{\text{gap}}^{2}/B_0}$, where $E_{\text{gap}}$ is the minimal gap. We neglect the possibility for transitions to the states $\vert\pm 1,-1,-1\rangle$ as a second order effect, occurring with probability of the order of $P_{1,\bm{k}}P_{2,\bm{k}}$.

\subsubsection{Analysis of the cooling dynamics}
\label{sec: analysis quantum channel}

 A single cycle of the cooling protocol, including the measurement of the ancilla spins, can be described as a map from the initial density matrix for the system (corresponding to the pseudospin $\mu_{\bm{k}}$) to the final one, i.e., the dynamics realize a quantum channel. 
The protocol consists of applying this quantum channel to the system multiple times, until a steady state is reached.

We number the four states participating in the dynamics $\left| 1,1,1\right\rangle$, $\frac{1}{\sqrt{2}}\left(\vert 1,1,-1\rangle+\vert 1,-1,1\rangle\right)$, $\left| -1, 1, 1\right\rangle$, $\frac{1}{\sqrt{2}}\left(\vert -1,1,-1\rangle+\vert -1,-1,1\rangle\right)$ by $0,1,2,3$ for brevity and denote the corresponding density matrix by
\begin{align}
\rho=\begin{pmatrix}\rho_{00} & \rho_{01} & \rho_{02} & \rho_{03}\\
\rho_{10} & \rho_{11} & \rho_{12} & \rho_{13}\\
\rho_{20} & \rho_{21} & \rho_{22} & \rho_{23}\\
\rho_{30} & \rho_{31} & \rho_{32} & \rho_{33}
\end{pmatrix}.
\end{align}

The final step of the protocol, in which $\tau_{\bm{J}}^z$ are measured and the signs of $B_{\bm{J}}$ are adjusted accordingly for the next cycle, is equivalent to a resetting operation taking $\eta_{\bm{k}}^z,\chi_{\bm{k}}^z\rightarrow1$ after the measurement. The action of this operation on the density matrix is given by
\begin{align}
\rho\underset{\text{reset}}{\rightarrow}\begin{pmatrix}\rho_{00}+\rho_{11} & 0 & \rho_{02}+\rho_{13} & 0\\
0 & 0 & 0 & 0\\
\rho_{20}+\rho_{31} & 0 & \rho_{22}+\rho_{33} & 0\\
0 & 0 & 0 & 0
\end{pmatrix}.
\end{align}

The unitary time evolution step acts as
\begin{align}
\rho\underset{\text{unitary}}{\rightarrow} \mathcal{U}\rho \mathcal{U}^{\dagger},\nonumber
\end{align}
where
\begin{align}
\mathcal{U}=\left(\begin{smallmatrix}\sqrt{\widetilde{P}_{2,\bm{k}}} &  &  & e^{i\theta_1}\sqrt{P_{2,\bm{k}}}\\
 & -\sqrt{P_{1,\bm{k}}} & e^{i\theta_2}\sqrt{\widetilde{P}_{1,\bm{k}}}\\
 & e^{i\phi_2}\sqrt{\widetilde{P}_{1,\bm{k}}} & e^{i\theta_2+i\phi_2}\sqrt{P_{1,\bm{k}}}\\
-e^{i\phi_1}\sqrt{P_{2,\bm{k}}} &  &  & e^{i\theta_1+i\phi_1}\sqrt{\widetilde{P}_{2,\bm{k}}}
\end{smallmatrix}\right),
\end{align}
where $P_{1,\bm{k}}$ and $P_{2,\bm{k}}$ are the transition probabilities between the different states (see Supplementary Note 1\ref{sec: solution three spin problem} and Supplementary Fig.~\ref{fig. 0d tfi spectrum vs time}b), and
we denote the complementary probabilities to $P_{n,\bm{k}}$ by $\widetilde{P}_{n,\bm{k}}=1-P_{n,\bm{k}}$. Finally, under the full protocol cycle consisting of unitary evolution followed by measurement and reset,
\begin{align}
&\begin{pmatrix}\rho'_{00}\\
\rho'_{02}\\
\rho'_{20}\\
\rho'_{22}
\end{pmatrix}=M\begin{pmatrix}\rho_{00}\\
\rho_{02}\\
\rho_{20}\\
\rho_{22}
\end{pmatrix},
\end{align}
where
\begin{align}
M=\begin{pmatrix}\widetilde{P}_{2,\bm{k}} &  &  & \widetilde{P}_{1,\bm{k}}\\
 & \sqrt{\widetilde{P}_{2,\bm{k}}P_{1,\bm{k}}} & -\sqrt{P_{2,\bm{k}}\widetilde{P}_{1,\bm{k}}}\\
 & -\sqrt{P_{2,\bm{k}}\widetilde{P}_{1,\bm{k}}} & \sqrt{P_{1,\bm{k}}\widetilde{P}_{2,\bm{k}}}\\
P_{2,\bm{k}} &  &  & P_{1,\bm{k}}
\end{pmatrix}.
\end{align}
Here, the complex phases
$e^{i\theta_1},e^{i\theta_2},e^{i\phi_1},e^{i\phi_2}$ cancel. If we take $P_{2,\bm{k}}\rightarrow0$, then the steady state of the protocol is the desired ground state ($\mu_{\bm{k}}^z=1$) since there is no probability
for any transition.
In general, the steady state of these dynamics is given by 
\begin{align}
M\begin{pmatrix}\frac{\widetilde{P}_{1,\bm{k}}}{\widetilde{P}_{1,\bm{k}}+P_{2,\bm{k}}}\\
0\\
0\\
\frac{P_{2,\bm{k}}}{\widetilde{P}_{1,\bm{k}}+P_{2,\bm{k}}}
\end{pmatrix}=\begin{pmatrix}\frac{\widetilde{P}_{1,\bm{k}}}{\widetilde{P}_{1,\bm{k}}+P_{2,\bm{k}}}\\
0\\
0\\
\frac{P_{2,\bm{k}}}{\widetilde{P}_{1,\bm{k}}+P_{2,\bm{k}}}
\end{pmatrix};
\end{align}
the steady state is a mixture of the ground state ($\mu_{\bm{k}}^z=1$) and the excited state ($\mu_{\bm{k}}^z=-1$). The probability to be in the ground state is given by
\begin{equation}
\label{eq:P_steady}    P_{{\rm steady}, \bm{k}} =  \frac{1-P_{1,\bm{k}}}{1-P_{1,\bm{k}}+P_{2,\bm{k}}}.
\end{equation}

We denote the eigenvalues of the quantum channel by $\lambda_{n,\bm{k}}$ (where $\lambda_{1,\bm{k}}=1$ corresponds to the steady state). 
The eigenvalues, $\lambda_{2,\bm{k}}$, $\lambda_{3,\bm{k}}$, and $\lambda_{4,\bm{k}}$ 
determine the rate in cycles at which the steady state is exponentially approached. These are given by
\begin{align}
\label{eq: lambda 1}
\lambda_{2,\bm{k}}=\sqrt{1-P_{2,\bm{k}}}\sqrt{P_{1,\bm{k}}}+\sqrt{P_{2,\bm{k}}}\sqrt{1-P_{1,\bm{k}}},
\end{align}
\begin{align}
\label{eq: lambda 2}
\lambda_{3,\bm{k}}=\sqrt{1-P_{2,\bm{k}}}\sqrt{P_{1,\bm{k}}}-\sqrt{P_{2,\bm{k}}}\sqrt{1-P_{1,\bm{k}}}, 
\end{align}
\begin{align}
\label{eq: lambda 3}
\lambda_{4,\bm{k}}=P_{1,\bm{k}}-P_{2,\bm{k}}.
\end{align}
The largest among these (i.e., the eigenvalue closest to 1) is $\lambda_{2,\bm{k}}$, and hence this eigenvalue controls the approach to the steady state: asymptotically, the deviation from the steady state scales with the number of cycles $n$ as $e^{-n/n_0}$, where $n_0 = 1/\ln(1/\lambda_{2,\bm{k}})$. Note that, in general, $\lambda_{2,\bm{k}}\le 1$. In the special case $P_{1,\bm{k}}=1-P_{2,\bm{k}}$, $\lambda_{2,\bm{k}}=1$ and the steady state becomes
degenerate. For any other value of $P_{1,\bm{k}}$, the system converges to the steady state.



Interestingly, Supplementary Eq.~(\ref{eq:P_steady}) shows that for $P_{2,\bm{k}}\rightarrow0$, the steady state is exactly the ground state, 
independent of the value of $P_{1,\bm{k}}$.
This is due to the fact that, for $P_{2,\bm{k}}=0$, there is no way to escape from the ground state once it is reached. However, the {\it rate of approach} to the ground state does depend on $P_{1,\bm{k}}$.

\bibliography{References}